\documentclass[conference]{IEEEtran}

\usepackage{amssymb}
\usepackage{amsmath}
\usepackage{cite}
\usepackage{xcolor}
\usepackage[ruled,vlined]{algorithm2e}
\usepackage{url}
\usepackage{graphicx}
\usepackage{gensymb}
\usepackage{caption}
\usepackage{subcaption}

\def\D{\mathbf{D}}
\def\d{\mathbf{d}}

\def\H{{\mathbf{H}}}
\def\h{{\mathbf{h}}}
\def\y{{\mathbf{y}}}
\def\Y{{\mathbf{Y}}}
\def\w{{\mathbf{w}}}
\def\P{{\mathbf{P}}}

\begin{document}
%
\title{DFRC with Improved Communication-Sensing Trade-off via Private Subcarrier Permutations and Pairing with Antennas}

\author{\IEEEauthorblockN{Zhaoyi Xu and Athina P. Petropulu}
\IEEEauthorblockA{Department of Electrical and Computer Engineering, 
Rutgers University\\
Piscataway, New Jersey 08854 \\
}
}

%


\maketitle

\begin{abstract}
Dual function radar communication (DFRC) systems can achieve significant improvements in spectrum efficiency, system complexity and energy efficiency, and are attracting a lot of attention for next generation wireless system design. This paper considers DFRC systems using MIMO radar with a sparse transmit array, transmitting  OFDM waveforms, and assigning  shared and private subcarriers to active transmit antennas.
Subcarrier sharing \textcolor{black}{allows antennas to modulate data symbols onto the same subcarriers and} enables high communication rate, while the use of private subcarriers trades-off communication rate for sensing performance \textcolor{black}{by enabling the formulation of a virtual array with larger aperture than the physical receive array}.
We propose to exploit the permutation of private subcarriers among the available subcarriers and the pairing between active antennas and private subcarriers to  recover some of the communication rate loss. Exploiting the  $1$-sparse property of private subcarriers, we also propose a low complexity algorithm to identify private subcarriers and detect the antenna-subcarrier pairing.
\end{abstract}


%
\IEEEpeerreviewmaketitle

{
  \renewcommand{\thefootnote}{}%
  \footnotetext[1]{This work was supported by NSF under grant ECCS-2033433}
}

\section{Introduction}
Next-generation wireless applications aim to achieve  unconstrained access to spectrum for radar and communications, thereby achieving high spectral efficiency.
This has given rise to a lot of interest in designing systems that can coexist in the spectrum while using different platforms \cite{Babaei2013SpectrumSharing,Wang2015SpectrumSharing, Li2016SpectrumSharing}, or to  Dual Function Radar Communication (DFRC) systems that perform sensing and communication from a single platform \cite{Fan2020}. The former class can work with existing systems but requires means for controlling the interference between the two systems, for example, via a control center \cite{Li2016SpectrumSharing}. On the other hand, 
DFRC systems require  new signaling designs, but  do not require interference control \textcolor{black}{between radar and communication}, and further, they offer reduced cost, lighter hardware, and lower power consumption. For those reasons, DFRC systems are of great interest to vehicular networks, WLAN indoor positioning, and unmanned aerial vehicle networks \cite{adas,jrc,5gvv,xu2020dfrc,Fan2020}.

Orthogonal frequency division multiplexing (OFDM) is a popular approach to achieve high communication rate and also deal with frequency selective fading. For those reasons, OFDM  has been widely used in modern communication applications
\cite{Nee2006ofdm802.11, 
Li2020PLCOFDM, 
Ergen2009Mobilebroadband}.
OFDM signals  have also been used for radar purposes, due to their ability to flexibly occupy the available spectral
resources \cite{ofdm}, and easily overcome frequency selective propagation effects.
OFDM waveforms for DFRC systems have been explored in several works \cite{ofdm,ofdmdfrc,Liu2017OFDMDFRC,Dokhanchi2018OFDMDFRC,liu2017multiobjective,liu2019robust,liu2020super}.  

In our previous work \cite{xu2020dfrc,xu2021wideband}, we proposed a DFRC system with OFDM waveforms, that uses a sparse MIMO radar and assigns shared and private 
subcarriers to the active transmit antennas.
Subcarrier sharing enables high communication rate by allowing multiple antennas to modulate data symbols onto the same subcarrier. The use of private subcarriers, i.e., subcarriers exclusive to one antenna, trades-off communication rate for sensing performance \textcolor{black}{by enabling the formulation of  a virtual array with larger aperture than the physical array. A sparse sensing problem can be formulated based on the virtual array and solved to provide refined target estimates.}
In the  SS-DFRC (subcarrier sharing - DFRC) system of \cite{xu2020dfrc,xu2021wideband} the transmit array contains a small number of radio frequency (RF) chains. 
In each channel use, each RF chain is  connected to one antenna, selected from a large array of antennas. 
The system communicates information via the transmitted OFDM data symbols\cite{xu2021wideband}, and also the pattern of active transmit antennas\cite{xu2020dfrc}, in a generalised spatial modulation (GSM) fashion. A multi-antenna communication receiver could estimate the indices of the active antennas and  the transmitted data symbols via sparse signal recovery (SSR) methods. 
\textcolor{black}{
However, \textcolor{black}{\cite{xu2021wideband}, follows a less computationally complex approach, by exploiting the private subcarriers. Namely, it first identifies the indices of the active antennas by looking at their corresponding private subcarriers \textcolor{black}{whose indices were assumed to be fixed and known to the communication receiver}.
Then,  focusing  on the active antennas, the transmitted symbols can be obtained via a least-square approach.
}
\textcolor{black}{The OFDM SS-DFRC approach of \cite{xu2020dfrc,xu2021wideband} can achieve gigabits per second communication rate, which is much higher than previous approaches.}}


\textcolor{black}{Exploiting the transmit array sparsity, SSR  can be used to identify those private subcarriers and estimate the transmitted data symbols on both shared and private subcarriers. However, this would require the application of  SSR  on all subcarriers, which  would introduce large computational cost and  delay. Furthermore, if the receive array has more antennas than the transmit array, SSR methods are no longer suitable. In the context of GSM, active antenna detection was studied in \cite{Wang2012generalised} via simple linear matrix operations, alas, including an exhaustive search over all the possible antenna patterns. As the number of possible patterns grows in a combinatorial fashion,  the approach of \cite{Wang2012generalised} is applicable only to systems with a small number of antennas.
}

Inspired by \cite{Wang2012generalised}, we propose a method that involves linear matrix operations and  rapidly identifies the private subcarriers. The proposed algorithm exploits the $1$-sparse property of private subcarriers  and a binary search plus projection approach to decide whether a certain subcarrier is shared or private, and also find the corresponding active antenna if it is a private subcarrier. Once the communication receiver has identified all private subcarriers and detected the active antenna indices, the estimation of the symbols on the shared subcarriers can be carrier out via  a matched filter that maximizes the SNR while satisfying the constraint of eliminating the data symbols from other active antennas. As shown in the experiment results, \textcolor{black}{compared with SSR method, the proposed algorithm can be more than $1,000$ times faster in identifying the private subcarriers and the corresponding active antennas when the number of receive antennas is less than the number of transmit antennas. Furthermore, the proposed algorithm is also feasible when the receive array has more antennas than the transmit array.
} 

\textcolor{black}{{The contribution of this paper is twofold. \textit{First},  aiming to  further increase the communication rate and thus recover the rate loss from the use of private subcarriers in OFDM SS-DFRC, we propose a novel approach for conveying information via the permutation of private subcarriers as well as the pairing of private subcarriers  with active antennas.}  \textcolor{black}{\textit{Second},  we propose a method that involves linear matrix operations and  rapidly identifies the private subcarriers and the corresponding active antennas.
Given the fact that OFDM waveforms usually have a large number of subcarriers, if the communication receiver can correctly and efficiently identify the private subcarriers and recover the pairing, the overall communication rate can  be significantly increased.
}
}

\section{The SS-DFRC system with private subcarriers}
Let us consider an SS-DFRC OFDM system with $N_x$ RF chains, which can be connected to a uniform linear array (ULA)  with $N_t$ transmit and $N_r$ radar receive antennas, spaced apart by $g_t$ and $g_r$, respectively. 
In each channel use, $N_x$ antennas out of the $N_t$ available ones are selected to transmit. Each active antenna is connected to an RF chain. 
We will denote by ${\mathcal{N}}$ the set of selected antenna indices.

\begin{figure}
    \centering
    \includegraphics[width = 6cm]{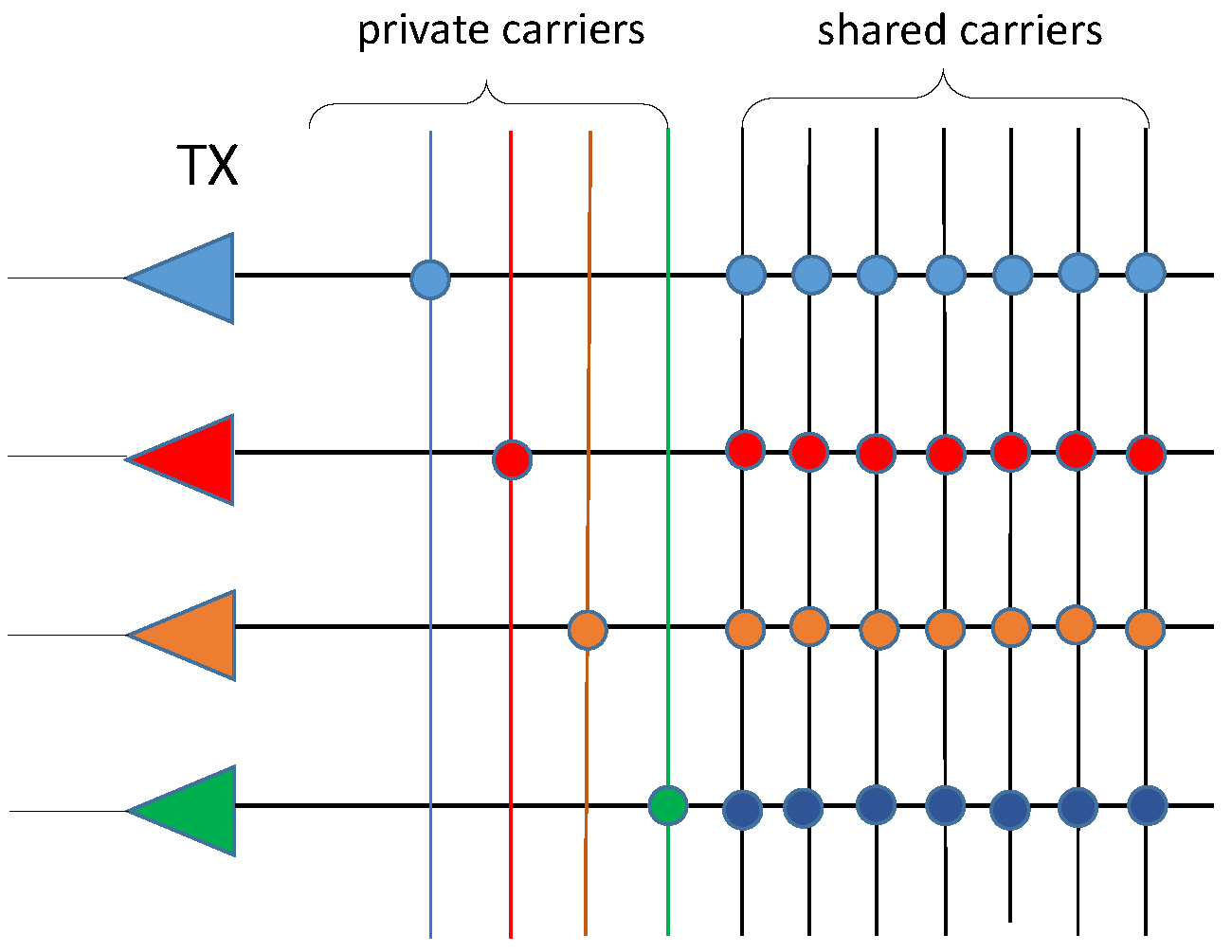}
    \caption{Private and shared subcarriers}
    \label{fig:subcarrier}
\end{figure}

The transmit waveforms are OFDM signals with $L$ subcarriers spaced by $\Delta f$, which are generated as follows.
The binary source data are divided into $N_x+1$ parallel streams,  among which $N_x$ streams are modulated 
and  then  distributed to the OFDM subcarriers assigned to each RF chain. The subcarriers can be shared or private;  shared subcarriers can be accessed by all RF chains. The remaining stream is conveyed in a  GSM fashion and will be discussed later.
\textcolor{black}{An illustration  of shared and private subcarriers is shown in Fig.~\ref{fig:subcarrier}.}

Each selected antenna applies an inverse discrete Fourier transform (IDFT) on the data symbols assigned to it, pre-appends a cyclic prefix (CP), converts the samples into an analog multicarrier signal and transmits it with carrier frequency $f_c$. 
That signal  will be referred to as  OFDM symbol, and has duration $T_p$. \textcolor{black}{In total there are $N_p$ OFDM symbols.}
Let $\D\in\mathbb{C}^{N_t\times L}$ denote a matrix that contains the data symbols to be transmitted during the $\mu$-th OFDM symbol, with elements $[\D]_{n,i}=d(n,i,\mu)$
i.e.,  the symbol  transmitted by the $n$-th antenna, on the $i$-th subcarrier, during  the $\mu$-th OFDM symbol. 
\textcolor{black}{The $i$-th column of $\D$, i.e., ${\mathbf{d}}_i$, contains the symbols transmitted by all antennas on subcarrier $i$, while the $j$-th row of $\D$ contains the symbols transmitted by the $j$-antenna on all subcarriers. Since only $N_x$ antennas are selected to transmit, the rows corresponding to not selected antennas will contain zeros.
Unless otherwise indicated, the use of the symbol matrix will refer to one OFDM symbol, thus, for notational simplicity, the dependence on $\mu$ is not shown in the notation $\D$. }

The complex envelope of the transmitted baseband signal on the $i$-th subcarrier due to the $n$-th antenna equals
\begin{align}
    x(n,i,t) = \sum_{\mu=0}^{N_p-1} d(n,i,\mu) e^{j2\pi i{\Delta f} t}rect(\frac{t-\mu T_p}{T_p}), \\ n \in \mathcal{N},\quad i = 0,1,\dots,L-1 \nonumber
    \label{eqform}
\end{align}
where $rect(t/T_p)$ denotes a rectangular pulse of duration $T_p$ and ${\Delta f}$ is the subcarrier spacing.

\textcolor{black}{The sensing functionality of the system is  detailed in \cite{xu2021wideband}. \textcolor{black}{In \cite{xu2021wideband}, the coupling of target parameters and data symbols can be addressed by first estimating the target angles based on the receive array, and then use those estimates to obtain range and Doppler estimates via cross-correlation operations. Using the private subcarriers, a virtual array  is constructed,  based on which a sparse sensing problem is formulated and solved to refine the angle estimates. For  details please refer to \cite{xu2021wideband}.}
In this paper we only focus on the communication component.}

\section{GSM on private subcarriers and pairing with active antennas}

Only a few antennas ($N_x<N_t$) are active in each OFDM symbol. \textcolor{black}{Each  active antenna is connected with one RF chain.} The pattern of active antennas along with the pairing between active antennas and private subcarriers can be used  to convey information and thus increase the communication rate.
Unlike our previous work \cite{xu2020dfrc}, here, the private subcarriers are randomly distributed, and the permutation of private subcarriers is also used to encode information.
Let the indices of $N_x$ active antennas be described by the set $\mathcal{N} = \{a_1,a_2,\dots,a_{N_x}\}$ where $a_i \in \{1,2,\dots,N_t\}$,  and the corresponding indices of private subcarriers be described by the set  $\mathcal{I} = {i_1,i_2,\dots,i_{N_x}} \in \{1,2,\dots,L\}$.

\textcolor{black}{Compared to shared subcarriers, each private subcarrier only transmits one data symbol. \textcolor{black}{The bit loss on each private subcarrier is $Q*(N_x-1)$, where $Q$ is the number of bits in each data symbol.}. Thus, $N_x$ private subcarriers lead to a total loss of $Q*(N_x-1)*N_x$ bits in each OFDM symbol However, this rate loss can be compensated by the permutation of private subcarriers and the pairing between active antennas. }
There are in total ${N_x \choose N_t}*P(N_x,L) $ combinations where $P(*,*)$ denotes permutation, thus $\lfloor\log_2{[{N_x \choose N_t}* P(N_x,L)]}\rfloor$ bits can be conveyed in each OFDM symbol, \textcolor{black}{where $\lfloor \cdot \rfloor$ denotes the floor function.}

Let us consider a communication receiver with $M$ antennas. Also, let $\Y\in \mathbb{C}^{M\times L}$ denote the received data symbol matrix in one OFDM symbol, and $\y_i$  the received data symbol on the $i$-th subcarrier. Also, let $\H  \in \mathbb{C}^{M\times N_t\times L}$ denote  the channel tensor, containing channels between all transmit receive antenna pairs and all subcarriers,   and $\H_i\in \mathbb{C}^{M\times N_t}$ denote one slice of $\H$, corresponding to the   channel matrix on the $i$-th subcarrier.
We assume here that the channels are  zero-mean, unit-variance  complex Gaussian distributed.
If  $i$ is the index of a shared subcarrier then it holds that
\begin{equation}
    \y_i = \H_i \d_i = \sum_{n=1}^{N_x} \h_{a_n i}d(a_n, i,\mu) + \w
    \label{eq:received_sym}
\end{equation}
\textcolor{black}{where $\d_i$ is the $i$-th column of symbol matrix $\D$,}
$\h_{n i}$ refers to the $n$-th column of matrix $\H_i$ and $\w$ denotes complex white Gaussian noise with variance $\sigma_c^2$. We assume that the channel matrix $\H$ is fully estimated at the receiver side, \textcolor{black}{which can be done by sending pilots in advance\cite{coleri2002channel}.}
\textcolor{black}{
If $i$ is the index of a private subcarrier,  then the  corresponding data symbol vector $\d_i$, $i\in\mathcal{I}$ is $1$-sparse, i.e., it has only one non-zero entry.}

If the communication receive array has fewer antennas than the radar
transmit array, i.e., $M < N_t$,
the sparse data symbol vector $\d_i$ can be recovered by finding the $\tilde{\d_i}$ with the minimum  $\ell_1$-norm that gives rise to ${\y}_i$, i.e.,  
\begin{equation}
    \begin{split}
        &\min \quad ||\tilde{\d_i}||_1\\
        &\ \text{s.t.} \quad ||\y_i - \H_i \tilde{\d}_i||_2^2 \leq M\sigma_c^2.
    \end{split}
    \label{l1norm1}
\end{equation}
From the sparse solution $\tilde{\d_i}$, one can directly obtain the estimated data symbols and the corresponding indices of active antennas.
However, the SSR method cannot decide whether the $i$-th subcarrier is shared or private until the calculation \textcolor{black}{of finding $\tilde{\d_i}$} 
has being done. Given the fact that the private subcarriers are randomly distributed and unknown to the receiver, applying SSR method on all subcarriers to identify them would take a large computation overhead and cause unacceptable delay.


If we neglect the noise, one can see that $\y_i$ in \eqref{eq:received_sym} is a linear combination of channel vectors corresponding to active antennas and belongs to the vector space $V =$ span$(\h_{a_1 i},\dots, \h_{a_{N_x} i})$.
Suppose that the vector spaces spanned by channel vectors corresponding to different active antenna indices sets are distinct, which can be satisfied if $M > N_t$. Upon finding the vector space that \textcolor{black}{has the minimal vertical distance} to $\y_i$, one could find the active antenna indices\cite{Wang2012generalised}.
However, such approach  would need exhaustive searches over all the possible antenna patterns, and thus can only be used in  systems with a small number of antennas.
For example, there are $2^{23}$ possibilities to select 
 $8$  out of $32$ antennas.  

\subsection{Identifying the private subcarriers via a binary search and projection approach}

Each private subcarrier only transmits one data symbol in each OFDM symbol, thus the corresponding received data symbol vector is naturally $1$-sparse. If we can  identify the positions of the sole data symbols on carriers that have already been identified as $1$-sparse, we can obtain the active antennas pattern.
Next,  we propose a method to detect the private subcarriers and the location of the sole data symbol in the transmitted symbol vector that is based on the well-known binary search algorithm and the projection idea of \cite{Wang2012generalised} but does not require exhaustive search.


Let us consider the $i$-th subcarrier and divide the  corresponding channel matrix into two sub-matrices, i.e.,
\begin{equation}
    \H_{i} = [\Tilde{\H}_{1}, \Tilde{\H}_{2}]
\end{equation}
where $\Tilde{\H}_{1}$ contains the first $\lfloor \frac{N}{2} \rfloor$ columns and  $\Tilde{\H}_{2}$ contains the remaining  columns of $\H _i$. 
\textcolor{black}{Since the transmitted data vectors on private subcarriers are $1$-sparse,}
if the $i$-th subcarrier is a private subcarrier, \textcolor{black}{the vertical distance from $\y_i$ to the projection of it on one subspace is just noise,  while the vertical distance to the projection on the other subspace is the channel vector modulated by data symbol and the noise.}
\textcolor{black}{Thus, by comparing the vertical distance from $\y_i$ to the two subspaces with the length of noise, we can know which subspace $\y_i$ lies in.}
Let us compute  the distance from the received symbol vector $\y_i$ to the subspaces spanned by the columns of $\Tilde{\H}_{1}$ and $\Tilde{\H}_{2}$. 
For  $\Tilde{\H}_{1}$ (and similarly for $\Tilde{\H}_{2}$) the distance  is $||(\mathbf{I}_M-\P_1)\y_i||$ where
\begin{equation}
   \P_1 =  \Tilde{\H}_{1}(\Tilde{\H}_{1}^H\Tilde{\H}_{1})^{-1}\Tilde{\H}_{1}^H
\end{equation}
and $\mathbf{I}_M$ represents an M by M identity matrix.

By iterating the above process  on the sub-matrix whose column  space contains $\y _i$, we will finally detect the sole channel vector corresponding to the active antenna.
If the $i$-th subcarrier is a shared subcarrier, in the middle of the process one would find that the received symbol vector lies in both subspaces.
By repeating the above  process on all the subcarriers, we can identify all the private subcarriers and the corresponding active antenna indices. 
The process  is described in Algorithm \ref{alg:private}, where $\epsilon = (M\sigma_c^2)^{\frac{1}{2}}$ \textcolor{black}{indicates the length of noise after projection.}
In order to guarantee the orthogonal between the column spaces of channel sub-matrices, $M>\frac{N_t}{2}$ should be satisfied. 

\begin{algorithm}[]
\SetAlgoVlined
\DontPrintSemicolon
Input: $\H_i$, $\y_i$, $i = 0,1,\dots,L-1$\;
\While{Number of columns in the input channel matrix $\neq$ 1}{
Step1: Divide the columns of input channel matrix into two groups and construct two sub-matrices $\Tilde{\H}_{1}$, $\Tilde{\H}_{2}$\;
Step2: Calculate the corresponding projection matrices $\P_1$ and $\P_2$\;
Step3:\; \eIf{$||(\mathbf{I}_M-\P_1)\y_i|| \leq \epsilon$ \& $||(\mathbf{I}_M-\P_2)\y_i||\leq \epsilon$}
{Subcarrier $i$ is a shared one,
Return;}
{\If{$||(\mathbf{I}_M-\P_1)\y_i|| \leq \epsilon$}
{Continue with $\Tilde{\H}_{1}$}
\If{$||(\mathbf{I}_M-\P_2)\y_i|| \leq \epsilon$}
{Continue with $\Tilde{\H}_{2}$}
Subcarrier $i$ is a shared one,
Return;
}
}
\caption{Fast private subcarrier identification and active antenna index estimation}
\label{alg:private}
\end{algorithm}

\textcolor{black}{Note that, the channel matrix can also be divided into more than $2$ sub-matrices, and apply the proposed algorithm. }

\subsection{Data symbol estimation via projection}
\label{sec:estimation}
Once we have estimated the active antenna indices, the corresponding data symbols of those active antennas can be estimated using projections as follows.

The received symbol vector on a shared subcarrier \eqref{eq:received_sym} now can be rewritten as
\begin{align}
      \y_i &= \sum_{n=1}^{N_x} \h_{a_n i}d(a_n, i,\mu) + \w \nonumber\\
      &= \h_{a_n i}d(a_n, i,\mu) + \sum_{n=1, i\neq n}^{N_x} \h_{a_n i}d(a_n, i,\mu) + \w.
      \label{6}
\end{align}
where the first term is the data symbol from the $n$-th active antenna and the seconds terms consists of data symbols from all other active antennas.
If $M > N_x$, we can assume that
the channel vectors $\h_{a_n i}$ for $i = 1,\dots, N_x$ are linear independent. 
Therefore, by  projecting $\y_i$ onto the orthogonal complement of the column space of $ \Tilde{\H}_{ni} = [\h_{a_1 i},\dots,\h_{a_{n-1} i},\h_{a_{n+1} i},\dots,\h_{a_{N_x} i}] $ we can
 eliminate the summation term in \eqref{6}, and get
 \begin{equation}
    \Tilde{\y}_{ni} = \Tilde{\P}_{ni}\y_i = \Tilde{\P}_{ni} \h_{a_n i}d(a_n, i,\mu) + \Tilde{\P}_{ni}\w
    \label{eq:7}
\end{equation}
 where
  $\Tilde{\P}_{ni}$ is the projection matrix corresponding to  the $n$-th active antenna
  and is equal to
\begin{equation}
    \Tilde{\P}_{ni}  = \mathbf{I}_M - \Tilde{\H}_{ni}(\Tilde{\H}^H_{ni}\Tilde{\H}_{ni})^{-1}\Tilde{\H}^H_{ni}.
\end{equation}

\textcolor{black}{On pre-multiplying \eqref{eq:7} by }  $\h^H_{a_n i}\Tilde{\P}^H_{ni}$, the first term becomes $||\Tilde{\P}_{ni} \h_{a_n i}||^2 d(a_n, i,\mu)$, which is a scaled version of the data symbol (the scalar is positive). 


\textcolor{black}{
 The above operation is equivalent to filtering that maximizes the output SNR while  eliminates interference from other active antennas. 
We should  mention that the above projection is equivalent to the solution of a least square problem.
}
Note that the conditions $M>N_x$ should be satisfied to guarantee the orthogonality between channel vectors corresponding to active antennas. Thus, $M>\max\{N_x,N/2\}$ should be satisfied.
Since $N_x < N_t$, the proposed algorithm works when $\max\{N_x,N/2\}< M< N_t$, and as shown in the  results section, it outperforms the SSR method in terms of lower runtime and also in terms of higher probability to successfully identify the private subcarriers and the corresponding active antennas.

\section{Numerical results}

Monte-Carlo experiments were conducted to validate the effectiveness of the proposed approach. The runtime (dashed line) and probability of identifying all the private subcarriers and their pairing with active antennas in one OFDM symbol (solid line) are plotted. After the estimation of active antennas, data symbol estimation (see  Sec.~\ref{sec:estimation}) is done via a projection-based method which is equivalent to a least-square method, and {the corresponding accuracy on data symbol estimation via least-square method was evaluated in
\cite{xu2021wideband},} 
and we omit the corresponding bit error rate result for simplicity. 

The data streams are randomly generated binary numbers,  modulated by quadrature phase shift keying (QPSK). The system parameters are shown in Table~\ref{table:system_parameters}.
\begin{table}[]
\caption{System Parameters}
\label{table:system_parameters}
\centering
\resizebox{60mm}{9mm}{
\begin{tabular}{ |c||c|c|  }
 \hline
 Parameter & Symbol & Value\\
 \hline
 Subcarrier spacing &   ${\Delta f} $  & 0.25MHz\\
 Duration of  OFDM symbol & $T_p$ & 5$\mu$s\\
 Number of subcarriers & $L$ & 64\\
 Total number of transmit antennas & $N_t$ & 32\\
 Number of activated antennas & $N_{x}$ & 6\\
 \hline
\end{tabular}}
\label{table:parameters}
\end{table}

As shown in Table~\ref{table:system_parameters}, $\lfloor\log_2{[ P(N_x,L)]}\rfloor = 23$ bits can be conveyed via the permutation of private subcarriers and their pairing with active antennas.
\textcolor{black}{
The use of private subcarriers results in a  rate loss of $11.44$ megabits/second. However,  via pairing and permutation of private subcarriers we can recover  up to $8.01$  megabits/second.
}
Here,  the  CP  time is included in $T_p$.

\begin{figure}[]
            \centering
            \includegraphics[width = 8.5cm]{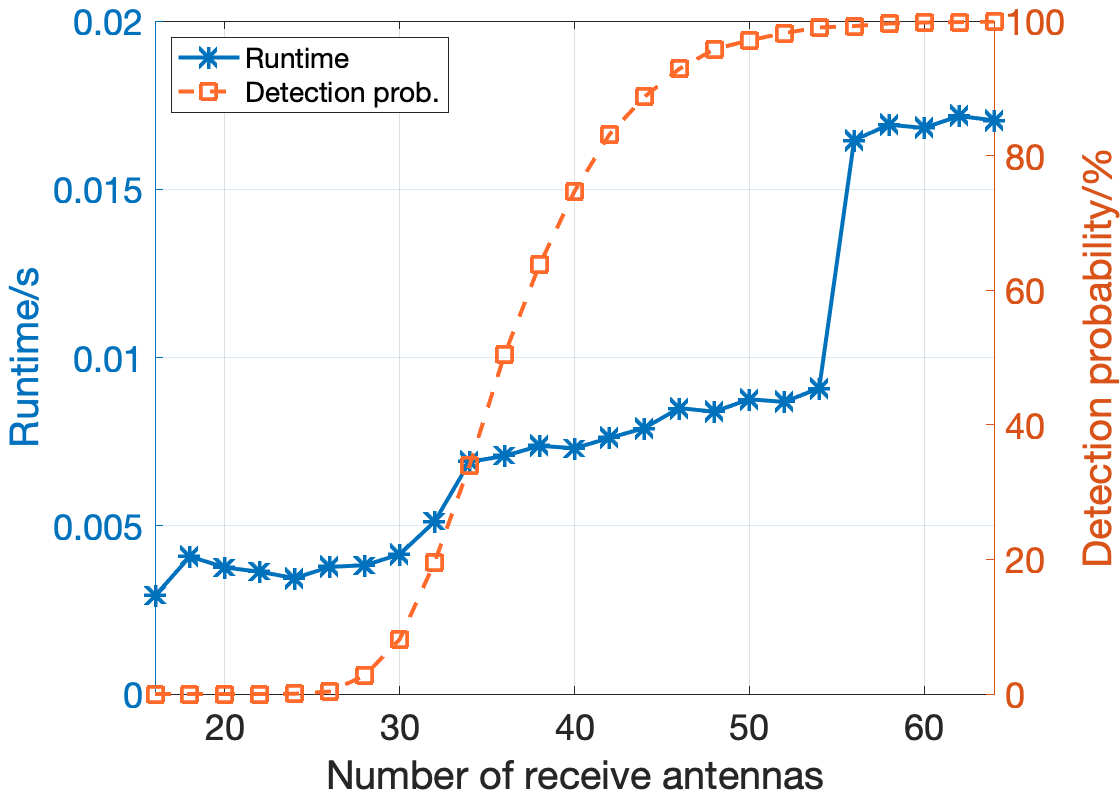}
            \caption{Runtime and detection probability with different numbers of receive antennas when $N_t=32$, $N_x=6$, SNR = $0$dB}
            \label{fig1}
        \end{figure}

In the first experiment, the performance of the proposed method for different numbers of receive antennas is shown in Fig.~\ref{fig1}. Here, the SNR is $0$dB and for each $M$, $10000$ experiments were repeated. The number of receive antennas increases from $16$ to $64$ with step of $2$.
As shown in Fig.~\ref{fig1}, \textcolor{black}{when the number of receive antennas increases,} the proposed method has a higher probability to identify the private subcarriers and the corresponding active antennas,  and also has a longer runtime. 
From the figure, we can observe that for $M$ between $48$ and $54$, a relative low runtime is achieved while the detection probability is close to $100\%$. 
\textcolor{black}{When $M$ is close to $N/2$, i.e., $16$, one can see from the figure that the proposed method fails to detect private subcarriers and the corresponding active antennas, since the subspaces of $\Tilde{\H_1}$ and $\Tilde{\H_2}$ are highly overlapped.} \textcolor{black}{Given that the proposed method involves matrix inversion operation, and based on the fact that in Matlab, the matrix inversion algorithm depends on the input matrix,   the runtime is not smooth and has a rapid increase at $M=56$.}
For the scenario of SNR = $0$dB considered, the proposed method is robust to noise and can carry out the detection in less than $0.02$ seconds. 

\begin{figure}[]
            \centering
            \includegraphics[width = 8.5cm]{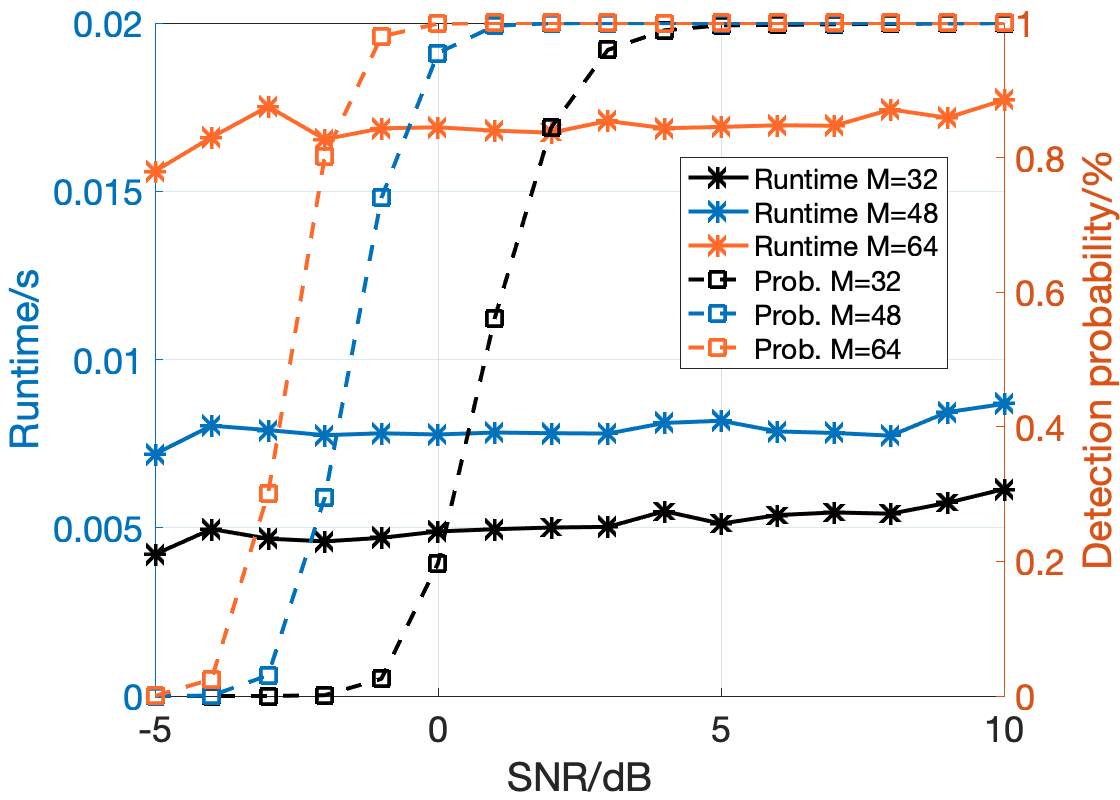}
            \caption{The performance of proposed algorithm with different SNR and different $M$, $N_t=32$, $N_x=6$.}
            \label{fig2}
        \end{figure}

The performance of the proposed algorithm with different SNR levels and three different numbers of receive antennas is shown in Fig.~\ref{fig2}, where each experiment configuration is repeated $10,000$ times.
The SNR was set to increase from $-5$dB to $10$dB, and $M$ was set to be $16$, $32$ and $64$.
From Fig.~\ref{fig2}, one can observe that $M=48$ is most efficient in terms of  runtime and detection probability, which is consistent with the observations made in Fig.~\ref{fig1}. Meanwhile, for $M=32$, the detection probability increases rapidly from $0.197$ to $0.96$ when the SNR increases from $0$dB to $3$dB, which suggests that if the noise level is low, one could use a small numbers of receive antennas to save in runtime while still having a high detection probability. 

\begin{figure}[]
            \centering
            \includegraphics[width = 8.5cm]{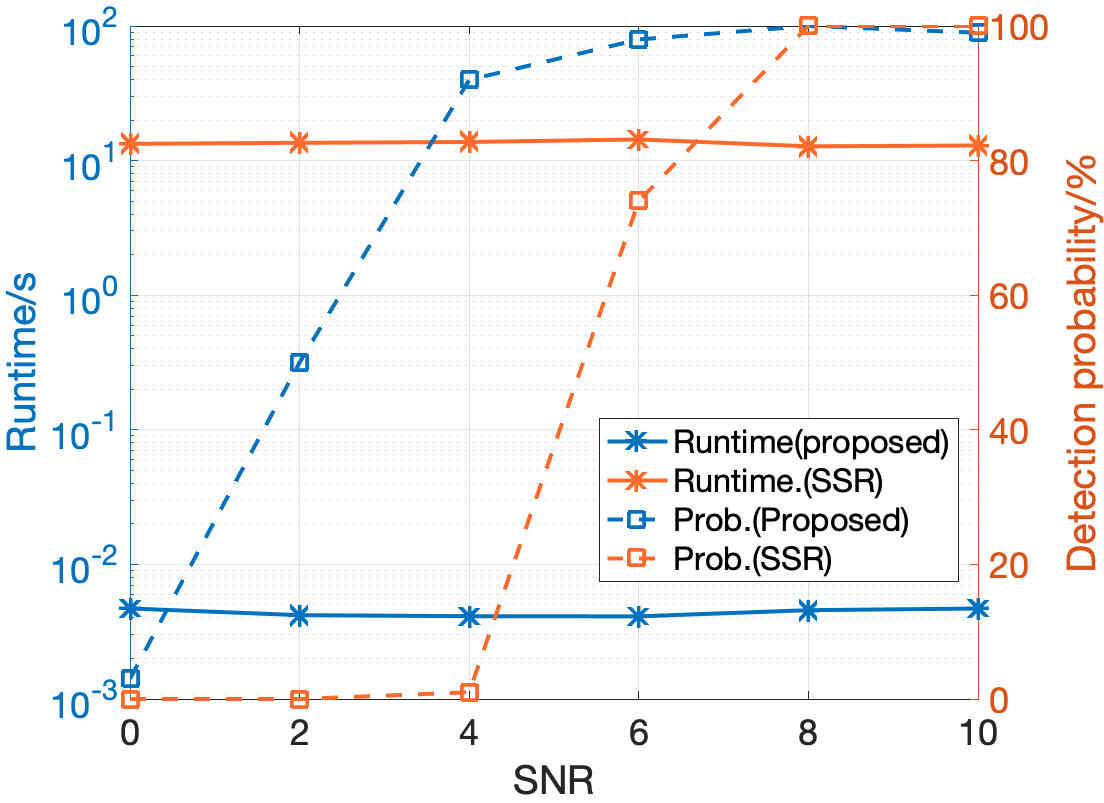}
            \vspace{-2mm}
            \caption{Comparison between the proposed algorithm and SSR method. $N_t=32$, $N_x=6$, $M= 28$}
            \label{fig3}
        \end{figure}

In order to further analyze the performance of the proposed algorithm when $M<N_t$ and compare it with the SSR method  used in our previous work, we plot Fig.~\ref{fig3}, where each experiment configuration is repeated $100$ times. The SSR method is implemented in Matlab and we used CVX, a package for specifying and solving convex programs \cite{cvx} to solve \eqref{l1norm1}.
In this experiment, the number of receive antennas is set to be $28$ which is smaller than the number of transmit antennas, so that SSR method can work. From the figure, one can clearly notice that the proposed method's runtime to detect all the private subcarriers is of the order of  tens of milliseconds, while that of the SSR's   is of the order of tens of seconds. 
\textcolor{black}{Furthermore, even through the subspaces of $\Tilde{\H_1}$ and $\Tilde{\H_2}$ are partially overlapped when $M<N_t$, the proposed method is still feasible and has higher probability to identify those private subcarriers. As shown in Fig.~\ref{fig3}, the detection probability is over $90\%$ when SNR is $4$dB.}

During the experiments, we found that the proposed algorithm performs better when we added a small offset $0.01*M$ to the threshold $\epsilon$ in Algorithm.~\ref{alg:private}.

\section{Conclusion}
In this paper, we have proposed to convey more information in an OFDM SS-DFRC system via the permutation of private subcarriers and the pairing between active antennas which renders megabits/second increment in communication rate. To enable a fast detection of private subcarriers and identifying the corresponding active antennas, based on the $1$-sparse property of private subcarriers, we have also proposed a low complexity algorithm which takes the idea of binary search and matrix projection. Through extensive experiments, we have validated the efficiency of the proposed algorithm which outperforms the SSR method.






\bibliographystyle{IEEEtran}
\bibliography{ref.bib}
%

\end{document}